## Chapter 8

# Distributed data analytics

*Richard Mortier[1] Hamed Haddadi[2]*
*Sandra Servia[3] and Liang Wang[4]*

Machine Learning (ML) techniques have begun to dominate data analytics applications and services. Recommendation systems are a key component of online service providers such as Amazon, Netflix and Spotify. The financial industry has adopted ML to harness large volumes of data in areas such as fraud detection, risk-management, and compliance. Deep Learning is the technology behind voice-based personal assistants [1], self-driving cars [2], automatic image processing [3], etc. Deployment of ML technologies onto cloud computing infrastructures has benefited numerous aspects of our daily life.

The advertising and associated online industries in particular have fuelled a rapid rise the in deployment of personal data collection and analytics tools. Traditionally, behavioural analytics relies on collecting vast amounts of data in centralised cloud infrastructure before using it to train machine learning models that allow user behaviour and preferences to be inferred. A contrasting approach, *distributed data analytics*, where code and models for training and inference are distributed to the places where data is collected, has been boosted by two recent, ongoing developments: (*i*) increased processing power and memory capacity available in user devices at the edge of the network, such as smartphones and home assistants; and (*ii*) increased sensitivity to the highly intrusive nature of many of these devices and services and the attendant demands for improved privacy.

Indeed, the potential for increased privacy is not the only benefit of distributing data analytics to the edges of the network: reducing the movement of large volumes of data can also improve energy efficiency, helping to ameliorate the ever increasing carbon footprint of our digital infrastructure, enabling much lower latency for service interactions than is possible when services are cloud-hosted. These approaches often introduce challenges in privacy, utility, and efficiency trade-offs, while having to ensure fruitful user engagement. We begin by discussing the motivations for distributing analytics (§1.1) and outlining the different approaches that have been taken (§1.2). We then expand on ways in which analytics can be distributed to the very edges of the network (§1.3), before presenting the Databox, a platform for supporting distributed analytics (§1.4). We continue by discussing personalising (§1.5) and scaling (§1.6) learning on such a platform, before concluding (§1.7).

[1] Cambridge University
[2] Imperial College London
[3] Cambridge University
[4] Cambridge University

## 1.1 Why distribute analytics?

Large-scale data collection from individuals is at the heart of many current Internet business models. Many current Internet services rely on inferences from models trained on user data. Data-driven solutions are now pervasive in areas such as advertising, smart cities and eHealth [4], [5]. Commonly, both the training and inference tasks are carried out using cloud resources fed by personal data collected at scale from users. Almost everything we do in our daily lives is tracked by some means or another. Although the careful analysis of these data can be highly beneficial for us as individuals and for the society in general, this approach usually entails invasion of privacy, a high price that progressively more people are not willing to pay [6]. Holding and using such large collections of personal data in the cloud creates privacy risks to the data subjects, but is currently required for users to benefit from such services.

Access to these data allow companies to train models from which to infer user behaviour and preferences, typically leveraging the generous computation resources available in the public cloud. Most current approaches to processing big data rely on centralised, usually cloud-hosted, storage. This makes it easy to conduct analyses of datasets but has drawbacks: (*i*) the regulatory climate evolves, making centralised storage riskier and costlier; (*ii*) mashing up inherently distributed datasets (e.g. those originating from different partners in a supply chain) is awkward when the algorithms to do so are designed to operate on centrally stored data; (*iii*) even simple data sharing is challenging due to uncertain commercial risks, hence the comparative lack of traction of 'open data' efforts such as Google Fusion Tables [7] compared to 'open APIs'; (*iv*) such data collection is increasingly pervasive and invasive, notwithstanding regulatory frameworks such as the EU's General Data Protection Regulation (GDPR) which attempt to restrain it.

To date most personal data were sensed through consumers' computers, smartphones or wearable devices such as smartwatches and smart wristbands. But nowadays smart technology is entering into our homes. We are heading towards an ecosystem where sooner or later, every device in our home will talk to an Amazon Echo, Google Home, or Apple HomeKit. Apart from controlling smart home appliances such as light bulbs and thermostats with our voice, these smart controllers for the entire home will be required to perform more complex tasks such as detecting how many people are in the house and who they are, recognising the activity they are performing or even telling us what to wear, or understanding our emotions [8].

The result is that user privacy is compromised, and this is becoming an increasing concern due to reporting of the ongoing stream of security breaches that result in malicious parties accessing such personal data. But such large-scale data collection causes privacy to be compromised even without security being breached. For example, consider wearable devices that report data they collect from in-built sensors, e.g., accelerometer traces and heart rate data, to the device manufacturer. The device might anonymise such data for the manufacturer to use in improving their models for recognising the user's current activity, an entirely legitimate and non-invasive practice. However, the manufacturer might fail to effectively anonymise these data and instead use them for other purposes such as determining mood, or even to sell to third-parties without the users' knowledge. It is not only data from wearables that creates such risks: web queries, article reads and searches, visits to shopping sites and browsing online catalogues are also indexed, analysed, and traded by thousands of tracking services in order to build preference models [9].

In these new scenarios, users are becoming progressively more aware of the privacy risks of sharing their voice, video or any other data sensed in their homes with the service providers, at the same time that these applications are demanding more accurate and personalised solutions. Sending personal data to the public cloud to perform these tasks seems no longer to be an acceptable solution, but solutions should take advantage of the resource capabilities of personal devices and bring the processing locally, where data resides.

Consider a home containing IoT devices, e.g., a smart energy meter. Occupants may be unwilling to release raw data to an external party for security, privacy, or bandwidth reasons, but might be willing to release certain summaries, especially if there is mutual benefit, e.g., from an external party sharing value generated by combining summaries from many users. The external party specifies which analyses each analytics engine should compute, and the occupants confirm they accept the exchange.

Larger organisations can also benefit from such a facility, by holding their own datasets and providing for analyses based on requests from external parties. Examples might include data tracking provenance and reliability of components in a complex supply chain where the organisation would control the sorts of summaries that can be computed and shared, e.g., allowing a third party to discover which supplier's components are problematic without having to reveal to them details of all current suppliers. Similarly, the National Digital Twin being pursued by the UK Government [10] can only feasibly be built by providing a federated infrastructure that enables many disparate sources of data owned by many organisations to contribute to a single whole. This saves the costs of transmitting and copying large datasets, reduces risks of bulk copying copyright or commercially sensitive datasets, and simplifies the contractual complexity of obtaining permissions from the various parties involved in data-mashup analyses. The same approach can even apply when privacy and regulatory issues are absent, e.g., decentralised processing is needed for live telemetry from remote vehicles with only a satellite uplink, or for ultra-high frequency monitoring of switches in a datacenter.

We are interested in alternative approaches where we reduce or remove the flow of user data to the cloud completely, instead moving computation to where the data already resides under the user's control [11]. This can mitigate risks of breach and misuse of data by simply avoiding it being collected at scale in the first place: attack incentives are reduced as the attacker must gain access to millions of devices to capture data for millions of users, rather than accessing a single cloud service. However, it presents challenges for the sorts of model learning processes required: how can such models be learnt without access to the users' personal data?

These approaches require pervasive support for accountability [12], to make uses of data legible to data subjects, owners, and controllers as well as to data processors, given that computations may be distributed to untrusted hosts. They also require development of novel underlying infrastructure in which computations can be efficiently distributed and results efficiently aggregated while exposing sufficient metadata about the operation of the system through interfaces designed for that purpose, so all users can be satisfied they have sufficient understanding of past, present, and future activities involving their data and analytics.

## 1.2 Approaches to distribution

Some analytics are easy to distribute. In the smart energy metering example, suppose an energy company wants to know the mean and variance of daily energy usage per customer. Each customer's analytics engine would compute and share its individual mean and variance. The data analyst at the energy company would aggregate these summary statistics, and compute the overall mean and variance. This could be an automatic weekly process, reporting the same statistics every week to underpin the energy company's dashboard. In this way each customer limits how much data she shares, and the energy company limits how much data it collects and stores. Ad hoc queries will be needed as well as dashboards. E.g., to answer the question "How did energy consumption change in the recent cold snap?" the data scientist might ask each customer to summarise their data differently according to weather data by region.

More advanced data science queries use iterative computation. Building a predictive model for how a customer's energy consumption varies over time might start, for example, with a rough predictive model, then ask each customer to compute a summary statistic about the prediction quality of this model, aggregate summaries to improve the model, and iterate. Such distributed computations are behind classic machine learning algorithms such as Google's PageRank. This pushes real-time distributed computation to its limits. Scaling this demands new techniques such as rapid iteration on small samples followed by slow iteration on large samples; using data from customers willing to share lots of their data to refine the questions to be asked of the rest of the population; making efficient use of summaries as they trickle in; settling for less accurate answers that can be computed from simpler summaries; or using randomly re-sampled representative data from each customer rather than actual traces.

### 1.1.1 Distributed analytics

Distributed machine learning (DML) arose as a solution to effectively use large computer clusters and highly parallel computational architectures to speed up the training of big models over the large amounts of data available nowadays [13]. Systems dealing with very large datasets have already had to handle the case where no single node can contain and process the entire dataset, but the dataset and/or the model to learn are parallelised among different machines, models are sequentially trained on each single machine, and some sort of synchronisation mechanism is applied to aggregate the parameters of the model to learn [14]–[17]. DML may also be a potential solution when the volume of the data is not the main issue, but the distribution occurs when different entities own different datasets which, if aggregated, would provide useful knowledge. However, the sensitivity of such data often prevents these entities from sharing their datasets, restricting access to only a small set of selected people as in the case of patients' medical records [18].

Gradient descent is the workhorse for many machine learning algorithms. Distributing it to run on big data is theoretically straightforward if the dataset is sharded 'horizontally' so each worker holds a few rows. This style of algorithm has driven the development of big data platforms, e.g., Spark and GraphLab [17], [19]. Refinements address unevenly distributed data [20], and high communication overhead [21]. Related methods for iterative optimisation have also been studied [21], [22].

Big data allows us to explore heterogeneity using high-dimensional models. Topological models like PageRank and nearest-neighbour have one parameter per data point, and distributed computation requires that we shard the data according to topology [23]. Deep neural networks can be distributed by sharding the parameters [21]. While appropriate in a datacenter, wide-area distribution remains an open problem.

Much modern statistics is based on MCMC algorithms for fitting large complex Bayesian models. These are computationally intensive and do not scale well to large datasets, which prompted work on distributed approaches using small random subsamples of data [24], along with studies of asymptotic correctness [25], [26]. Methods often rely on consensus techniques [27], expectation propagation [28] or stochastic optimization [29], [30]. But this is not settled science: even in simple real-valued parameter estimation there are issues of bias [31]; and when data is missing (e.g., as users choose not to share their data, or because the statistical model involves hidden variables) the compute burden increases drastically [32].

### 1.1.2 Federated & hybrid learning

Recently, Federated Learning (FL) [33] has become a popular paradigm in machine learning. In FL, a global model is trained using continuous, async updates received from users' devices in a privacy-preserving and decentralised manner. This approach helps users train large-scale model using their private data, while minimising the privacy risks. FL addresses a number of challenges traditionally associated with large-scale models, including device availability, complexities with local data distribution (e.g., time zone or the number of users within a region), reliance on device connectivity, interrupted execution across devices with varying availability; and challenges with device storage and processing resources. One of the main design criteria is that the process should not interfere with a user's device usage experience and should be done when the device is idle.

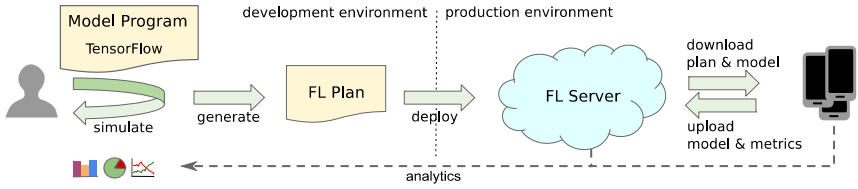

Figure 8.1 Google's Federated Learning framework [33].

Recently there has been a rise in hybrid models for distributed analytics, where parts of large Deep Neural Network (DNN) models run locally on the device, providing local feature extraction on the raw data, then only sending these features to the cloud for further classification and analysis [34], [35]. These works often work rely on the assumption that service providers can release publicly verifiable feature extractor modules based on an initial training set. The user's device then performs a minimalistic analysis and extracts private features from the data and sends it to the service provider (i.e., the cloud) for subsequent analysis. The private features are then analysed in the cloud and the result yields back to the user. The fundamental challenge in using this framework is the design of the feature extractor modules that removes sensitive information accurately, without major impact on scalability by imposing heavy computational requirements on the user's device.

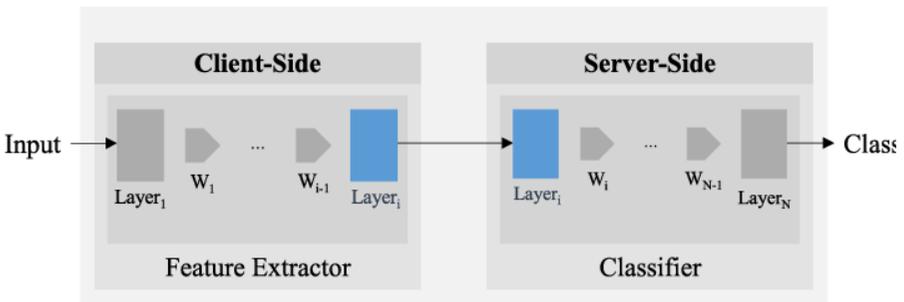

Figure 8.2 Hybrid Deep Learning frameworks [34], [35].

### 1.1.3  Personalised Learning

Contrary to distributed or federated learning, personalised learning does not seek to learn a global model from sensitive data from multiple parties, but to learn a personalised model for each individual party by building on a model learnt from a relatively small set of other parties but without requiring access to their raw data. Personal learning models [36] are similar to personal recommender systems [37], but generalise the solution to any learning algorithm. This approach takes advantage of transfer learning [38] to achieve better performance than algorithms trained using only local data, particularly in those common situations where local data is a scarce resource. This solution brings most of the data processing to where the data resides and not the other way around, exactly as the edge computing paradigm calls for [34], [39]. Recent popular machine learning frameworks provide libraries for running complex deep learning inferences on local devices such as smartphones [40]–[42]. While in these works models are previously trained in an offline manner, researchers have demonstrated that both the inference and the local retraining can be performed locally on low-power device in a timely manner [36].

### 1.1.4 Securing data

Approaches such as homomorphic encryption allow user data to be encrypted, protecting against unintended release of such data, while still being amenable to data processing. This affords users better privacy – their data cannot be used arbitrarily – while allowing data processors to collect and use such data in cloud computing environments. However, current practical techniques limit the forms of computation that can be supported.

Lately, Trusted Executions environments (TEEs) such as Arm TrustZone or Intel Software Guard Extensions (SGX) have been used for providing attestation and trust in models and data being executed on the cloud and on user devices. Using these approaches, a secure area in memory can be used for securely storing sensitive data (e.g., a fingerprint scan or face features), or providing model privacy and defence against adversarial attacks such as Membership Inference Attacks [43].

While numerous scalable distributed computation platforms exist, e.g., [19], [44], they do not target the wide-area deployment or multi-tenant hosting of computation we require, and neither are they designed and built to support accountability. Databox [11] supports accountability but presumes small-scale localised personal data.

Glimmers [45] explores using Intel SGX at the client to create a trusted party that can validate accuracy of client contributions before adding to an aggregate, without revealing the details of the contribution further. Distributed, federated provenance has been studied [46], [47]. Tango [48] provides a replicated, in-memory data structure backed by a shared log for recording such data.

## 1.3 Analytics at the edge

Increased ubiquity of sensing via mobile and IoT devices has caused a surge in personal data generation and use. Alongside this surge, concerns over privacy, trust, and security are becoming increasingly important as different stakeholders attempt to take advantage of such rich data resources: occurrences of breaches of privacy are rising at alarming rates [49]. Tensions in the collection and use of personal data, between the benefits to various analytics applications, the privacy consequences and security risks, and the regulatory complexities of aggregating and processing data in the cloud are a significant barrier to innovation in this space. We have previously proposed that these topics, and the shortcomings of current approaches in this space, are the concern of a new – or at least, newly focused – discipline, Human-Data Interaction (HDI) [50].

In our view the core technical problem in this space is how to build networked services that enable individuals to manage their personal data so that they can permit other parties to make use of it while retaining personal control over such uses and understanding the implications of any data release. As digital data may be copied infinitely at negligible marginal cost and without loss of fidelity, the current common approach of centralising unencrypted personal data into a cloud-hosted service such as Google or Facebook is fundamentally flawed in this regard. Once data is given up to such a service, the data subject can only exercise control over it to the extent that the cloud provider allows them, and they have only social means (e.g., negative publicity campaigns, local per-jurisdiction regulation) to ensure such controls are provided.[5]

Personal data collection for profiling and mining users' interests and relationships is the basis on which online platforms such as Facebook and Google and services such as Apple's Siri and Microsoft's Cortana operate. However, such data collection and profiling exposes the user to privacy leakage even when these communities are anonymous [51]. Simultaneously, these cloud-based services can have only a partial view of each data subject's digital footprint, resulting in inaccuracies and systemic biases in the data they hold, and leading to ever more aggressive data collection strategies.

Ever since Edward Snowden revealed the secret large-scale government-level surveillance programmes, social awareness of privacy and personal data is quickly arising. The Internet Trends 2016 report [52] points out that, according to its survey, 45% respondents feel more worried about their online privacy than one year ago, and 74% have limited their online activity because of privacy concerns.

Building privacy, trust and security into the evolving digital ecosystem is thus broadly recognized as a key societal challenge. Regulatory activities in the US [53], Europe [54] and Japan [55] are complemented by industry initiatives that seek to rebalance the "crisis in trust" [12] occasioned by widespread personal data harvesting. All parties agree that increased accountability and control are key to this challenge. *Accountability* seeks not only to strengthen compliance but also to make the emerging ecosystem more transparent to consumers, while *control* seeks to empower consumers and provide them with the means of actively exercising choice.

Many online service providers who collect large-scale data from users are prone to data breaches. Users rely on the promises from big companies to keep their data private and safe [56]. However, these providers are not infallible. Yahoo, Tumblr, and Ashley Madison (an online dating service for married people) offer only a few high-profile examples of breaches in recent years [57].

There have been many responses to these challenges. Making databases private and secure is one of the solutions. Cloud Kotta [58] introduces a cloud-based framework that enables secure management and analysis of large, and potentially sensitive, datasets. It claims to ensure secure data storage that can only be accessed by authorised users. Joshi et al [59] present an ontology so that big data analytics consumers can write data privacy policies using formal policy languages and build automated systems for compliance validation. Cuzzocrea et al [60] presents a short review of existing research efforts along this line.

---

[5] We follow standard legal terminology and refer to individuals about whom data is collected from *sources* such as IoT sensors or online social network accounts as *data subjects*, and organisations wishing to process data as *data processors*.

Hiding data alone is not enough. The model itself can also reveal private information. For example, Fredrikson et al [61] showed that access to the face recognition model enables recovery of recognisable images. Abadi et al [62] develops a method that can prevent such model inversion attacks even by a strong adversary who has full knowledge of the training mechanism and access to the model parameters. Similarly, Papernot et al [63] propose a "teacher-student" approach based on knowledge aggregation and transfer so as to hide the models trained on sensitive data in a black box.

Several privacy-preserving analytical solutions have been proposed to guarantee the confidentiality of personal data while extracting useful information [18], [64]–[66]. Prominent among them are those that build on Dwork's *differential privacy* framework [62], [67]–[70], which formalises the idea that a *query* over a sensitive database should not reveal whether any one person is included in the dataset [71]. In the case of machine learning, the idea is that a differentially-private model should not reveal whether data from any one person were used to train the model. Most of these techniques for differentially-private machine learning are usually based on adding noise during the training, which leads to a challenging trade-off between accuracy and privacy.

Shokri and Shmatikov [72] and McMahan *et al.* [73] propose solutions where multiple parties jointly learn a neural-network model for a given objective by sharing their learning parameters, but without sharing their input datasets. A different approach is proposed by Hamm *et al.* [74] and Papernot *et al.* [63], where privacy-preserving models are learned locally from disjoint datasets, and then combined on a privacy-preserving fashion. However, the privacy guarantees of some of these solutions have recently been called into question [75]. Data privacy when data is sharded 'vertically,' so each worker holds a few columns, has been studied as a problem of secure communication using garbled circuits. Low-dimensional linear models can be fitted [76], but the extension to richer algorithms and to high-dimensional modelling is an open problem.

Instead of hiding data or models, some suggest the user should choose which part of data to upload. Xu et al [77] propose a mechanism to allow users to clean their data before uploading them to process. It allows for prediction of the desired information, while hiding confidential information that client want to keep private. RAPPOR [78] enables collecting statistics data from end-user in privacy-preserving crowdsourcing.

Although numerous mechanisms supporting privacy preserving analytics, marketing and advertising have been proposed, e.g., recent studies on analysing network traces using differential privacy [79] and accessing databases while respecting privacy [80], [81], no operational system exists that also gives others visibility into statistics and trends [82]–[84]. Rieffel *et al.* [85] propose cryptographic, hierarchical access to data for processing aggregate statistics without decrypting personal data. However this method still requires collection of individual data items and complex yet critical management of many cryptographic keys. Privacy-aware centralised methods such as homomorphic encryption [86] are yet to be deployed in a commercial or consumer system. While these methods are likely to be important in the future, they are not enough alone: they cannot provide accountability and control in isolation.

However, all of the aforementioned work focus on the traditional cloud side solution. Users' data are still collected to central server for processing, which are prone to issues such as increased service response latency, communication cost, and single point failure. Numerous recent and current projects and startups have responded to specific problems of aggressive data collection by the online advertising industry[6] through more traditional means. These typically involve production of services called variously *Personal Data Stores*, *Personal Information Management System*, *Vendor Relationship Management*, and similar; examples include Mydex [87] and openPDS [88]. They allow the subject to retain ownership[7] of their data and provide it to third parties on demand [89], which offers some degree of accountability and control but only insofar as the service provider can be trusted.

One solution is to deploy ML services on edge devices. Moving services from cloud to users' edge devices can keep the data private, and effectively reduce the communication cost and response latency. Some research has begun to emerge that aims to solve accompanied challenges. They recognise that mobile devices cannot afford to support most of today's intelligent systems because of the large amount of computation resource and memory required. As a result, many current end-side services only support simple ML models. These solutions mostly focus on reducing model size [90]–[92]. For example, Neurosurgeon [92], partitions a DNN into two parts, half on edge devices and the other half on cloud, reducing total latency and energy consumption. A key observation is that, in a DNN, output size of each node decreases from front-end to back-end, while the change of computation latency is the opposite.

Recently computation on edge and mobile devices has gained rapid growth, such as personal data analytics in home [11], Intel's Movidius Neural Compute Stick [93] is a tiny deep learning device that one can use to accelerate AI programming and DNN inference application deployment at the edge. Edge computing is also boosting content distribution by supporting peering and caching [94]. Huawei has identified speed and responsiveness of native AI processing on mobile devices as the key to a new era in smartphone innovation [95].

Moving ML analytics from cloud to edge devices faces many challenges. One widely recognised challenge is that, compared with resource-rich computing clusters, edge and mobile devices only have quite limited computation power and working memory. To accommodate heavy ML computation on edge devices, one solution is to train suitable small models to do inference on mobile devices [96]. This method leads to unsatisfactory accuracy and user experience. Some techniques [90], [91], [97] are proposed to enhance this method.

To reduce the memory and disk usage of speech recognition application, Lei et al [97] uses a compressed n-gram language model to do on-the-fly model rescoring. Chen et al [90] presents HashNets, a network architecture that can reduce the redundancy of neural network models to decrease model sizes, while keeping little impact on prediction accuracy. MobileNets [91] from Google reduces model size by a different technique: factorisation of the convolution operation.

---

[6] http://www.technologyreview.com/view/530741/the-murky-world-of-third-party-web-tracking/

[7] Simple notions of ownership are problematic given the inherently social nature of even personal data, and are dealt with elsewhere in this book but particularly Chapters 2 and 4.

## 1.4 Databox, a platform for edge analytics

Our response is to provide technical means to assist the data subject in managing access to their data by others. Perhaps the most extreme expression of this approach to date is the Databox, an open-source personal networked device augmented by cloud-hosted services that collates, curates, and mediates access to our personal data, under the data subject's control [11]. It gathers data from local and remote sources, from online social networks to IoT sensors; provides for data subjects to inspect data gathered from their data sources, and to effect actuation via IoT devices and similar; enables data processors to discover and request access to subjects with sources of interest; and it supports running applications to provide data processors with specific, limited, logged access to subjects' data.

Databox sits within an ecosystem of networked devices and associated services enabling individuals to manage their data, and to provide other parties with controlled access to their data. Composed of a set of service instances, realised as Docker-managed containers in our current prototype, it supports placing these instances in different locations, from a physical device in the subject's home, to the public cloud, to future envisioned edge-network hosting resources such as smart lampposts and cell-towers.

Databox not only benefits data subjects by providing a regulated and privacy-enhanced communication mechanism between data subjects and data processors. Acting as an agent on behalf of the data subject, it can support queries over high-resolution personal data that would be difficult for a single company to obtain, permitting richer, more accurate data analytics. It also helps avoid the risks of data breach associated with collecting and curating large, personal datasets which malicious actors have significant incentives to attack and steal.

We envisage that all Databox applications will involve some software component running within the Databox. Specifically, applications provide derived stores to which external parties (whether data processing organisations or a browser operated by the data subject) can connect. These applications provide accountable entities through which the data subject can ascribe to a data processor behaviour involving use of their data. We envisage two main routes to installation of these components, resulting from successful negotiation between the data subject and processor causing the processor to be given access to the subject's data: subject-driven and processor-driven.

**Subject-driven**. This model is strongly analogous to current app store models. Each app store is a service to which subjects can connect to discover applications they might wish to install. Apps advertise the data stores that they will consume, the frequency at which they will access those stores, and the computation they will carry out on data from those stores. Apps will be validated and verified by the app store operators, and rated by the community. They will have only limited, approved external communications between their own store and an authenticated client.

**Processor-driven**. This model inverts the interaction, enabling data processors to discover cohorts of data subjects who have data available to meet the processor's needs.

Independent of the discovery model used, applications may either be limited to 1:1 interactions, or may necessarily involve a larger cohort of subjects making data available. In the former case, the output of the computation is consumed in isolation, either by the data subject or the data processor. In the latter, there is an inherent need for the application to function as a distributed system, with communication taking place between instances as the computation makes progress. This latter case is considerably more complicated so we discuss it briefly next.

Three key challenges of this sort of application present themselves: **scale**, **heterogeneity**, and **dynamics**. These challenges arise due to the fundamental characteristics of the algorithms and deployment environments envisaged. Data processors might use machine learning and model generation algorithms that default to serial computation or, at best, execute in the controlled, high bandwidth, low latency environment of a datacenter. Scaling them across (potentially) millions of Databoxes moves most of these algorithms outside their normal operating regions. The computation resources on which they will be hosted will vary in capacity and connectivity, making scheduling and synchronisation of ongoing computations between instances considerably more complex, Finally, physical Databox instances are likely to have variable and unreliable connectivity which, when coupled with the envisaged scale, almost guarantees that the entire cohort will never be simultaneously available.

Approaches to addressing these challenges that we are exploring include the use of techniques such as delay-tolerant querying, introduced in the Seaweed database [98], where metadata statistics are incorporated into the decisions taken by the system as to when to wait for data to become available and when to give up and return (with appropriate indications) a potentially inaccurate answer; and more flexible control over synchronisation barriers than permitted by Bulk Synchronous Parallel operation (e.g., Stale Synchronous Parallel Parameter Server [16]).

It is possible that other factors inherent in these data and the deployment of Databoxes may also mitigate against some of these problems. For example, the distributed computations may well be highly localised and so might be loosely coupled and require minimal coordination and exchange of data. Coupled with use of aggregation in the computation graph, this might mitigate unreliability and scale, while also providing natural means to support privacy-preserving aggregation.

While it addresses concerns of privacy and ethics of these data, it does not try simply to *prevent* all such analysis and use by third-parties as not all of this activity is harmful [99], [100]. Rather, it seeks to afford users the possibility to find personal equilibria with sharing and use of their data: simply preventing all access to personal data would fail to take advantage of the many potential benefits that sharing data can produce, whether immediate financial rewards, social benefits through, e.g., participation in friendship groups, or broad societal benefits from the ability to participate in large-scale studies in, e.g., human mobility and activity for use in urban planning, or mental and physical health measures used to set healthcare norms.

## 1.5 Personalised learning at the edge

The approach of sending all users' personal data to the cloud for processing, is one extreme of a spectrum whose other extreme would be to train a model for a specific user using only that user's data. For some applications, e.g., activity recognition, it has been shown that a model trained solely using data from the individual concerned provides more accurate predictions for that individual than a model trained using data from other individuals [101]. At the same time, this solution offers more privacy to the user as all computation, for both training and inference, can be done locally on the device [102]. However, this approach leads to substantial interactional overheads as training the model will likely require each user to label a significant amount of data by hand before they can obtain accurate inferences.

An alternative is an inversion of the hybrid approach described in (§1.1.2): split computation between the cloud and the users' personal devices by (*i*) training a *shared model* in the cloud using data from a small (relative to the population) set of users, (*ii*) distributing this shared model to users' personal devices, where (*iii*) it can be used locally to generate inferences and (*iv*) it can be retrained using locally-stored personal data to become a *personal model*, specialised for the user in question [36].

In more detail, start by training a *shared* model, $M_S$, to recognise the activity that the user is performing using data sensed with his smartphone's built-in sensors. This *batch learning* is done on a remote server in the cloud using available public data, $d_p$. In the event of not having sufficient public data available for this task, data can be previously gathered from a set of users that have agreed to share their personal data perhaps by providing them with suitable incentives.

The user $u$ then obtains the *shared* model from the remote server. With every new sample or group of samples gathered from the smartphone's sensors, the activity that the user is performing is locally inferred using this model. In order to allow for more accurate inferences, the user is prompted to "validate" the results by reporting the activity they were performing. The new labelled data so gathered are then used for locally retraining the model, resulting in a new *personal* model, $M_P$.

This approach has been evaluated using (*i*) a neural network to recognise users' activity on the *WISDM* dataset [103] and (*ii*) the Latent Dirichlet Algorithm (LDA) [104] to identify topics in the Wikipedia and NIPS datasets [105], [106]. In both cases the model resulting from local retraining of an initial model learnt from a small set of users performs with higher accuracy than either the initial model alone or a model trained using only data from the specific user of interest.

We have briefly presented the software architecture for the Databox, a hybrid locally- and cloud-hosted system for personal data management (§1.4). We now sketch a system architecture that might be used to implement personalised learning using a device such as the Databox. This is divided into two parts: (*i*) residing in the cloud, the first part is responsible for constructing a *shared* model using batch learning; and (*ii*) residing on each individual user's device, the second part tunes the model from the first part using the locally available data, resulting in a *personal* model.

We identify five components in this architecture:

1. The **batch training module** resides in the cloud, and is responsible for training a *shared* model as the starting point using public, or private but shared, datasets that it also maintains. As this component may need to support multiple applications, it will provide a collection of different machine learning algorithms to build various needed models. It may also need to perform more traditional, large scale processing, but can easily be built using modern data processing frameworks designed for datacenters such as Mllib [107] or GraphLab [17].
2. The **distribution module** resides on users' devices and is responsible for obtaining the *shared* model and maintaining it locally. In the case of very large-scale deployments, standard content distribution or even peer-to-peer techniques could be used to alleviate load on the cloud service.
3. The **personalisation module** builds a *personal model* by refining the model parameters of the shared model using the personal data available on the user's device. This module will also require a repository of different learning algorithms, but the nature of personal computational devices means that there will be greater resource constraints applied to the performance and efficiency of the algorithm implementation.
4. The **communication module** handles all the communications between peers or those between an individual node and the server. Nodes can register themselves with the server, on top of which we can implement more sophisticated membership management.
5. The **inference module** provides a service at the client to respond to model queries, using the most refined model available.

In our implementation, we rely on several existing software libraries to provide the more mundane of these functions, e.g., ZeroMQ [108] satisfies most of the requirements of the communication and model distribution modules, and so we do not discuss these further here.

There are many toolkits, e.g., *theano* [109] and *scikit-learn* [110], that provide a rich set of machine learning algorithms for use in the batch training and personalisation modules. However, in the case of the latter, we must balance convenience with performance considerations due to the resource-constrained nature of these devices. In light of this, we use a more recent library, Owl [111], [112], to generate more compact and efficient native code on a range of platforms, and the source code can be obtained from its Github repository.[8]

We briefly summarise the workflow we envisage using activity recognition as an example.

1. When the user activates the device for the first time, the device contacts the server and registers itself in order to join the system. The device notices there is no local data for building the model, and sends a request to the server to obtain the *shared* model.
2. After processing the registration, the server receives the download request. The *shared* model has been trained using a initial dataset collected in a suitably ethical and trustworthy way, e.g., with informed consent, appropriate compensation, and properly anonymised. The server can either approve the download request, or return a list of peers from whom the requesting user can retrieve the model.

---

[8] https://github.com/owlbarn/owl

3. Having obtained the *shared* model, the device can start processing inference requests. At the same time, the device continuously collects user's personal data, in this case, their accelerometer traces. Once enough local data is collected, the personalisation phase starts, refining the shared model to create a *personal* model.
4. After the personal model has been built, the system uses it to serve requests, and continues to refine it as more personal data is collected.

The above system can suffer the attacks and consequences of malicious users. There are several potential attacks against any learning system [113], [114]. Here we focus on how privacy and causative attacks might affect our system. On a privacy attack the adversary obtains information from the learner, compromising the secrecy or privacy of the system's users. The aim of a causative attack is on altering the parameters of the target model by manipulating the training dataset. An example of this type of attacks are poisoning attacks, where an attacker may poison the training data by injecting carefully designed samples to eventually compromise the whole learning process. The target model then updates itself with the poisoned data and gradually compromises. Below we describe the potential effects of these attacks in our system.

### 1.1.5 Privacy attacks

Our solution guarantees the confidentiality of users' data (potential users) given that their devices are not compromised, since their personal data never leave their devices. Since both the data and the personal model reside on the user's device, attacks such as model inversion [70] –where an attacker, given the model and some auxiliary information about the user, can determine some user's raw data; and membership query [115], where, given a data record and black-box access to a model, an adversary could determine if the record was in the model's training dataset, cannot affect our users. However, we cannot assure the confidentiality of the data, neither robustness against these attacks, for those users that have freely agreed to share their data in the same way as the big corporations are not doing so with their customers data.

For many applications we envisage and describe in the introduction, such as those based on object recognition or those that work with textual data, there is already a large amount of data freely available on the Internet with which to build the shared model, and whose confidentiality does not need to be guaranteed. On the other hand, for applications such as face or speaker recognition, techniques based on differentially private training [63], [72]–[74] could be applied in order to, a priori, guarantee the confidentiality of the volunteers' data. On the contrary, the training of the personal model for the final users happens locally on their devices so that neither their data nor their personal model leave their devices, and its confidentiality is guaranteed by the security offered by their device, security that is out of the scope of the methodology proposed here.

### 1.1.6 Poisoning attacks

We envisage two different points or steps in our system that adversaries might wish to attack: when building the shared model in a remote server in the public cloud using public data available or *shared* by a group of volunteers, and when personalising the model by local retraining in the user's device (*personalisation*). In the case of a poisoning attack to our proposed methodology, the shared model can be corrupted by malicious volunteers poisoning the data with fake samples. However, during the local retraining, if the adversary wishes to corrupt the personal model, he needs to gain access to the local device of the user to poison the data and fool the model. Poisoning the data to train the personal model needs the attacker to gain access to the local device of the user.

Some schemes have been proposed to conduct poisoning attacks against SVMs [116], [117], but we have barely seen any work about poisoning attacks against neural networks. Our goal is not on how to design the best poisoning attack to achieve a given output or to avoid being detected, but on the effects of poisoning data into our model. Therefore we consider a dumb adversary that randomly alters labels into our training set without any other goal than misclassifying samples. In the following, we explore the effects of adding corrupted samples to the data used to train the shared model in the supervised learning task. Specifically, we simulate a scenario where part of the data used for training the *shared* model is corrupted. That is, a scenario where one or several volunteers intentionally alter the labels of their samples in the training set, and explore the effect that different amounts of data corrupted in the training set of the *shared* model cause in the *personal* model of the user.

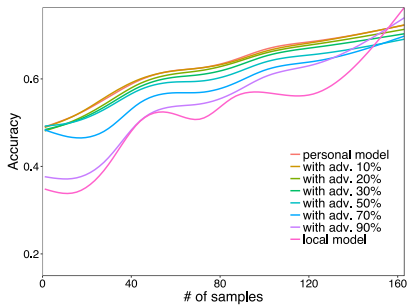
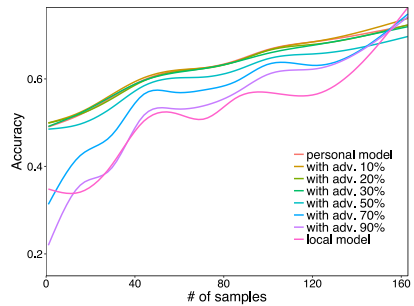

*(a)* *Accuracy of the personal model for different percentages of corrupted samples in the training set of the shared model ("dumb" adversary).*

*(b)* *Accuracy of the personal model for different percentages of corrupted samples in the training set of shared model ("smarter" adversary).*

Figure 8.3: Accuracy of the personal model in the face of dumb and not-so-dumb adversaries

To this aim, we corrupted different samples of the training set (10%, 20%, 30%, 50%, 70% and 90%). In order to generate each corrupted sample, we consider the original features of the sample to be corrupted, but assigning it a random activity. This random activity label was selected by a random generator with equal probability for each activity (walking, standing, etc.). Figure 3 represents the accuracy, in average, of the *personal* model for different percentages of corrupted samples in the training set of the *shared* model. We observe that the more corrupted samples used to build the *shared* model, the less improvement we get when starting to train the *personal* model. More interestingly, when less than 50% of the samples are corrupted, the accuracy obtained with the *personal* model is similar to the one when no samples are poisoned, especially when few samples are used to train the *local* model. Moreover, when the 90% of the samples are poisoned, the accuracy of the *personal* model is not worse than the *local* model.

Now we consider a "smarter" adversary that has access to other volunteers' data and therefore can obtain the distribution of labels in the dataset. In order to minimise the chances of being discovered, this adversary corrupts its samples by adding random labels with the probability of each label being the same as the one in the original data. Figure 3b shows the results when the labels corrupted follow the same distribution than the original sample ("smarter" adversary). In this case, we observe similar results as when poisoning the samples completely random. One difference is that when few samples of the concerned user are used to train the *personal* model, the system behaves worse than the *local* model, especially when the percentage of corrupted samples is large. But, when using more samples, the system behaves better than the local one and similar to the system under the effects of the "dumb" adversary. In conclusion, our system is quite robust against poisoning attacks against the *shared* model.

## 1.6 Extreme scaling at the edge

In contrast to a highly reliable and homogeneous datacenter context, the edge computing model espoused by systems such as Databox assume a distributed system consisting of a large number (tens of thousands) of heterogeneous nodes distributed over a wide geographical area, e.g., different cities. Their network connections are unreliable compared to a datacenter network, and the bandwidth capacity is variable. The nodes are not static but rather join and leave the system over time giving rise to non-negligible churn.

| System | Synchronisation | Barrier Method |
|---|---|---|
| MapReduce [44] | Map must complete before reducing | BSP |
| Spark [19] | Aggregate updates after task completion | BSP |
| Pregel [118] | Superstep model | BSP |
| Hogwild! [119] | ASP with system-level delay bounds | ASP, SSP |
| Parameter Servers [120] | Swappable synchronisation method | BSP, ASP, SSP |
| Cyclic Delay [121] | Updates delayed by up to $N$ 1 steps | SSP |
| Yahoo! LDA [122] | Checkpoints | SSP, ASP |
| Owl+Actor[111] | Swappable synchronisation method | BSP, ASP, SSP, PSP |

*Table 8.1 Classification of the synchronisation methods used by different systems.*

Each node holds a local dataset and, even though nodes may query each other, we do not assume any specific information sharing between nodes or between a node and a centralised server. Exemplar data analytics algorithms that might run across the nodes include Stochastic Gradient Descent (SGD) as it is one of the few core algorithms in many machine learning (ML) and deep neural network (DNN) algorithms. Synchronisation of model updates is achieved using one of several synchronisation control mechanisms. Table 8.1 summarises the synchronisation control used in different machine learning systems: how the different nodes participating in the computation coordinate to ensure good progress while maintaining accuracy.

*Bounded Synchronous Parallel (BSP)*

BSP is a deterministic scheme where workers perform a computation phase followed by a synchronisation/communication phase where they exchange updates [123]. The method ensures that all workers are on the same iteration of a computation by preventing any worker from proceeding to the next step until all can. Furthermore, the effects of the current computation are not made visible to other workers until the barrier has been passed. Provided the data and model of a distributed algorithm have been suitably scheduled, BSP programs are often serializable — that is, they are equivalent to sequential computations. This means that the correctness guarantees of the serial program are often realisable making BSP the strongest barrier control method [16]. Unfortunately, BSP does have a disadvantage. As workers must wait for others to finish, the presence of *stragglers*, workers which require more time to complete a step due to random and unpredictable factors [123], limit the computation efficiency to that of the slowest machine. This leads to a dramatic reduction in performance. Overall, BSP tends to offer high computation accuracy but suffers from poor efficiency in unfavourable environments.

*Asynchronous Parallel (ASP)*

ASP takes the opposite approach to BSP, allowing computations to execute as fast as possible by running workers completely asynchronously. In homogeneous environments (e.g. datacenters), wherein the workers have similar configurations, ASP enables fast convergence because it permits the highest iteration throughputs. Typically, $P$-fold speed-ups can be achieved [123] by adding more computation/storage/bandwidth resources. However, such asynchrony causes delayed updates: updates calculated on an old model state which should have been applied earlier but were not. Applying them introduces noise and error into the computation. Consequently, ASP suffers from decreased iteration quality and may even diverge in unfavourable environments. Overall, ASP offers excellent speed-ups in convergence but has a greater risk of diverging especially in a heterogeneous context.

*Stale Synchronous Parallel (SSP)*

SSP is a bounded-asynchronous model which can be viewed as a relaxation of BSP. Rather than requiring all workers to be on the same iteration, the system decides if a worker may proceed based on how far behind the slowest worker is, i.e. a pre-defined bounded staleness. Specifically, a worker which is more than $s$ iterations behind the fastest worker is considered too slow. If such a worker is present, the system pauses faster workers until the straggler catches up. This $s$ is known as the staleness parameter. More formally, each machine keeps an iteration counter, $c$, which it updates whenever it completes an iteration. Each worker also maintains a local view of the model state. After each iteration, a worker commits updates, i.e., $\Delta$, which the system then sends to other workers, along with the worker's updated counter. The bounding of clock differences through the staleness parameter means that the local model cannot contain updates older than $c - s - 1$ iterations. This limits the potential error. Note that systems typically enforce a read-my-writes consistency model.

The staleness parameter allows SSP to provide deterministic convergence guarantees [16], [123], [124]. Note that SSP is a generalisation of BSP: setting $s = 0$ yields the BSP method, whilst setting $s = \infty$ produces ASP. Overall, SSP offers a good compromise between fully deterministic BSP and fully asynchronous ASP [16], despite the fact that the central server still needs to maintain the global state to guarantee its determinism nature.

*Probabilistic Synchronous Parallel (PSP)*

PSP is a new barrier control technique suitable for data analytic applications deployed in large and unreliable distributed system. When used to build, for example, an SGD implementation, it effectively improves both the convergence speed and scalability of the SGD algorithm compared to BSP, SSP, and ASP. PSP introduces a new system primitive *sampling* that can be composed with existing barrier controls to derive fully distributed solutions. A full theoretical analysis of PSP is available [112] showing that it can provide probabilistic convergence guarantee as a function of sample size. PSP achieves better trade-off than existing solutions, i.e., iterate faster than SSP and with more accurate update. Both evaluation and theoretical results indicate that even quite small sample sizes achieve most of its benefits.

Practically all iterative learning algorithms are stateful. Both model and nodes' states need to be stored somewhere in order to coordinate nodes to make progress in a training process. Regarding the storage location of the model and nodes' states, there are four possible combinations as below (**states** in the list refer to the nodes' states specifically):

1. **[centralised model, centralised states]**: the central server is responsible for both synchronising the barrier and updating the model parameters,
   e.g., MapReduce [44] and Parameter Server [120] fall into this category.
2. **[centralised model, distributed states]**: the central server is responsible for updating the model only. The nodes coordinate among themselves to synchronise on barriers in a distributed way. P2P engine falls into this category.
3. **[distributed model, centralised states]**: this combination in practice is rare because it is hard to justify its benefits. We do not consider it further.
4. **[distributed model, distributed states]**: both model updates and barrier synchronisation are performed in a distributed fashion. A model can be divided into multiple chunks and distributed among different nodes.

Any of BSP, ASP, SSP, and PSP can be used in case 1 but only ASP and PSP can be used for cases 2 and 4. With PSP, the sever for maintaining the model can become "stateless" since it does not have to possess the global knowledge of the network. In case 2 particularly, the server takes the role of a stream server, continuously receiving and dispatching model updates, significantly simplifying the design of various system components.

Most previous designs tightly couple model update and barrier synchronisation but, by decoupling these two components using the *sampling* primitive, we can improve scalability without significant degradation of convergence. We present some simple simulation results here; for more detail see Wang et al [112]. For simplicity we assume every node holds the same amount of independent identically-distributed data. This evaluation covers both centralised and distributed scenarios. In the former, the central server applies the *sampling* primitive and the PSP implementation is as trivial as counting at the central server as it has global knowledge of the states of all nodes. In the distributed scenario, each individual node performs *sampling* locally whenever they need to make the decision to cross a synchronisation barrier e.g., to apply an update. We use the Owl library for all the numerical functions needed in the evaluation [111].

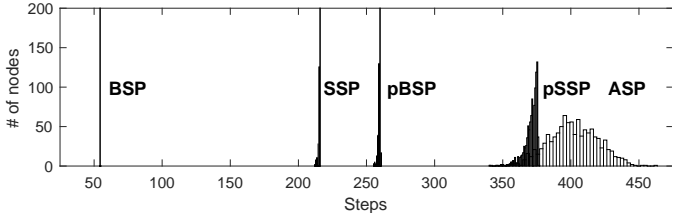

*(a)    Progress distribution in steps*

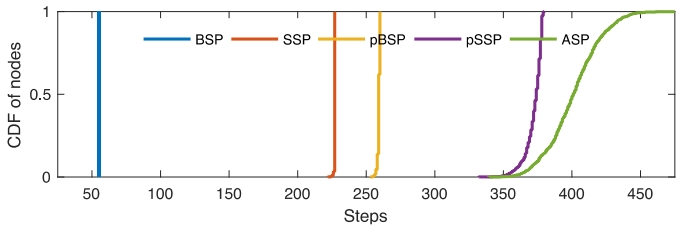

*(b)    CDF of nodes as a function of progress. No node maintains global state.*

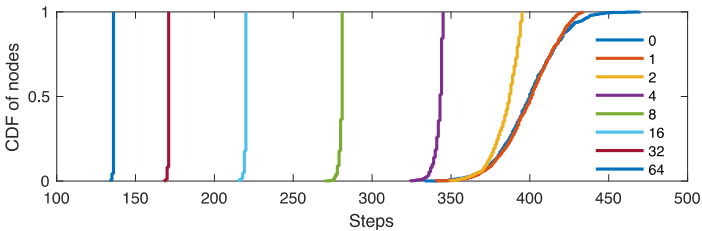

*(c)    pBSP parameterised by different sample sizes, from 0 to 64. Increasing the sample size decreases spread, shifting the curves to the left repli- cating behaviour from the most lenient (ASP) to the most strict (BSP).*

*Figure 8.4   SGD (stochastic gradient descendent) using five different barrier control strategies. Probabilistic Synchronous Parallel achieves a good trade-off between efficiency and accuracy.*

Figure 8.4 shows the results of evaluating PSP by simulating five different barrier control strategies for 40 seconds on a network of 1000 nodes running SGD algorithm. We use the parameter server engine to learn a linear model of 1000 parameters and each node uses a sample of 1% of the total number of nodes in the system unless otherwise specified.

Figure 8.4a plots the progress in steps of all nodes after the 40 simulated seconds. As expected, the most strict BSP leads to the slowest but most tightly clustered step distribution, while ASP is the fastest but most spread due to no synchronisation at all. SSP allows certain staleness (4 in our experiment) and sits between BSP and ASP. pBSP and pSSP are the probabilistic versions of BSP and SSP respectively, and further improve the iteration efficiency while limiting dispersion.

For the same experiment, Figure 8.4b plots the CDF of nodes as a function of their progress in steps. ASP has the widest spread due to its unsynchronised nature. Figure 8.4c focuses on pBSP synchronisation control with various parametrisation. In the experiment, we vary the sample size from 0 to 64. As we increase the sample size step by step, the curves start shifting from right to the left with tighter and tighter spread, indicating less variance in nodes' progress. With sample size 0, pBSP exhibits exactly the same behaviour as that of ASP; with increased sample size, pBSP starts becoming more similar to SSP and BSP with tighter requirements on synchronisation. pBSP of sample size 16 behaves very close to SSP regarding its progress rate.

Finally, we note that with a very small sample size, even just one or two, pBSP can already effectively synchronise most of the nodes compared to ASP but with dramatically smaller communication overheads. The tail caused by stragglers can be further trimmed by using larger sample size.

## 1.7 Conclusion

In this chapter we have covered the basic requirements and examples of distributed and decentralised analytics. A large number of these applications focus on optimisations between data utility and granularity, resources on the edge (memory, processing, battery), and privacy/security tradeoffs. Research in this space is providing novel methods for efficient and privacy-preserving methods for data collection, with the ability to provide instantaneous feedback and interaction with the user.

## 1.8 Acknowledgements

This work was supported by the Engineering and Physical Sciences Research Council [grant number(s) EP/N028260/1 and EP/N028260/2; EP/N028422/1; EP/G065802/1; EP/M02315X/1; EP/M001636/1; EP/R045178/1], and by the European Union Framework Programme 7 2007–2013 grant number 611001.